\documentclass[pra, 12 pt]{revtex4}
\usepackage[english]{babel}
\usepackage{amsmath}
\usepackage{epsfig}

\begin{document}


\title{\bf Finite Size Atom in the Hartree-Fock Approximation:\\ New Substance Quasiparticle}
\author{V.B. Bobrov, S.A. Trigger}
\address{Joint\, Institute\, for\, High\, Temperatures, Russian\, Academy\,
of\, Sciences, Izhorskaia St., 13, Bd. 2. Moscow\, 125412, Russia;\\
emails:\, vic5907@mail.ru,\;satron@mail.ru}

\begin{abstract}
It is shown that, in the self-consistent quantum statistical
Hartree-Fock approximation, the number of electronic states
localized on one nucleus is finite. This result is obtained on the
basis of the general electron-nuclear model of matter and provides
convergence of the atomic statistical sum and finiteness of the
"atom" size. In general approach the characteristic size of the
"atom" is a function of density and temperature. However, it is
shown, that in a wide range of thermodynamic parameters, for
relatively low temperatures, characteristic orbits and electron
energy eigenvalues are independent of density and temperature. In
this case, the sizes of the orbits are of order of the Bohr radius
which is a minimal characteristic size in the system for typical
parameters of plasma with atomic states.

PACS number(s): 31.15.-p, 51.30.+i, 52.25.Kn, 52.27.Gr

\end{abstract}

\maketitle

\section{Introduction}

Real matter (gases, liquids, solids, plasma) is constructed from
nuclei and electrons interacting with each other according to the
Coulomb law [1]. It would be very desirable, based on the known
nucleus charge as an initial problem parameter and, based only on
the Coulomb law, to construct a scheme for calculating macroscopic
properties of different media. The consistent approach to
implement such a scheme is based on the methods of the diagram
technique of the perturbation theory [1,2]. The main difficulties
in the realization of this approach are caused by the necessity of
the simultaneous consideration of both "localized" states of
electrons and nuclei ("atoms", "molecules", etc.), whose
description requires accurate summation of rows of the
perturbation theory, and "delocalized" states which are quite
successfully described within the lowest orders of the
perturbation theory [1,2]. As a result, "localized" states for
weakly nonideal plasma are considered within the second virial
coefficient [1-4]. In this case, there arises the known problem of
the convergence of the statistical sum of the hydrogen atom, which
is directly related to its size [2]. A large number of works are
devoted to the solution of this problem, where various procedures
and physical mechanisms of "cutoff" of the statistical sum of the
atom were proposed, including the papers [5-12] which have already
become classical (see [1-3] and references therein for more
details). In this case, the following circumstance is very
surprising. As is known [13,14], the self-consistent Hartree-Fock
approximation and its various generalizations are basic in
calculations of the structure and energy spectra of atoms and
molecules in quantum chemistry. Despite the known statement that
the self-consistent Hartree-Fock approximation is the best
single-particle approximation in quantum statistics [15], it is
almost not used in the study of thermodynamic properties of
Coulomb systems, where "localized" states should be taken into
account [1]. This paper is devoted to the solution of the problem
of convergence of the statistical sum of the hydrogen atom and its
size finiteness, based on the self-consistent quantum-statistical
Hartree-Fock approximation for the Coulomb system.

\section{Two-component electron-nuclear plasma: adiabatic approximation and classification of the electron states}

We consider the two-component Coulomb system (CS) in volume $V$ at
temperature $T$ within the canonical ensemble consisting of
electrons (subscript $e$) and nuclei (subscript $c$), taking into
account the quasineutrality condition $\sum_{a=e, c} z_aen_a^{(0)}
=0$. Here $n_a^{(0)} = \langle N_a \rangle^{(can)}/V$ is the
average density of the number of particles; $N_a$ is the operator
of the total number of particles of type $a$, characterized by
mass $m_a$, the average density of the number of particles $n_a$,
and charge $z_ae$; angle brackets $\langle \dots \rangle^ {(can)}$
mean averaging over the canonical ensemble. For such a CS, the
Hamiltonian $H^{CS} =H_{ee}+H_{cc} + H_{ec}$ in the secondary
quantization representation is written as [1]
\begin{eqnarray}
H_{aa}=- \frac{\hbar^2}{2m_a}\int \psi_a^+({\bf r})\nabla^2 \psi_a
({\bf r}) d {\bf r}+ V_{aa}, \;\;\; H_{ec}= \int u_{ec} ({\bf
r}_1-{\bf r}_2)N_e ({\bf r}_1)N_c ({\bf r}_2) d{\bf r}_1 d{\bf
r}_2, \label{F1}
\end{eqnarray}
\begin{eqnarray}
V_{aa}= \frac{1}{2} \int  u_{aa}({\bf r}_1-{\bf r}_2)\psi_a^+({\bf
r}_1) \psi_a^+({\bf r}_2) \psi_a({\bf r}_2) \psi_a({\bf r}_1) d
{\bf r}_1 d {\bf r}_2, \qquad u_{aa}( r)= \frac{z_az_be^2}{r} ,
\label{F2}
\end{eqnarray}

Here $u_{aa}(r)$ is the potential of the Coulomb interaction of
particles of types $a$ and $b$, $\psi_a^+({\bf r})$ and
$\psi_a({\bf r})$ are the field creation and annihilation
operators, $N_a({\bf r}) = \psi_a^+({\bf r})\psi_a({\bf r})$ is
the density operator of the number of particles of type $a$,
which, in the coordinate representation, is written as $N_a({\bf
r}) = \sum_{i=1}^{N_a}\delta({\bf r-R}_i^a)$. The values ${\bf
R}_i^a$ are the coordinates of particles of type $a$. We emphasize
that the total number of particles $N_a$ of each type in the
canonical ensemble is specified and is the C-number, $N_a = \int
N_a({\bf r}) d{\bf r} = \langle N_a \rangle^{(can)}$. The free
energy $F^{CS}$ of the considered quasi-neutral CS consisting of
the set number of electrons $N_e$ and nuclei $N_c$ in volume $V$
at temperature $T$, $F^{CS} =-T\ln Sp\left \{\exp \left(-\frac
{H^{CS}}{T}\right)\right\}$, in the adiabatic approximation for
nuclei, is written as
\begin{eqnarray}
F^{CS} \cong -T \ln Sp_c \left\{ \exp
\left(\frac{F_{ec}-H_{cc}}{T}\right) \right\}, \qquad F_{ec}=-T
\ln Sp_e \left\{ \exp \left(-\frac {H_{ee}+H_{ec}}{T}\right)
\right\}. \label{F3}
\end{eqnarray}

The quantity $F_{ec}$ (3) is the free energy of the subsystem of
electrons being in the external field of stationary nuclei. In
this case, the operator of the density of the number of nuclei
$n_c({\bf r})$ in the Hamiltonian $H_{ec}$ (1) is the C-number. At
this stage, when considering the subsystem of electrons in the
external field of stationary nuclei, we proceed from the
equivalence of the large canonical and canonical ensembles, i.e.,
we suppose that the equality [16]
\begin{eqnarray}
F_{ec}= \Omega_{ec}+\mu_e \langle N_e \rangle ^{(GE)}, \qquad
\Omega_{ec}=-T \ln Sp_e\left\{ \exp \left[- \frac
{H_{ee}+H_{ec}-\mu_e N_e}{T}\right] \right\}, \label{F4}
\end{eqnarray}
for the free energy $F_{ec}$ (3) is valid. In (4) the function
$\Omega_{ec}$ is the thermodynamic potential of the subsystem of
electrons with the chemical potential $\mu_e$ in the external
field of stationary nuclei, whose value depends on the nucleus
coordinates $\{{\bf R}_i^c\}$ as the external parameters, $\langle
N_e \rangle^{(GE)}$ is the average number of electrons as a
function of temperature $T$, volume $V$, and chemical potential
$\mu_e$. In this case, the quasineutrality condition takes the
form $\langle N_e \rangle^{(GE)} =z_cN_c$ and is used to determine
the chemical potential $\mu_e$ of electrons.

Then, to describe the field operators $\psi_a^+({\bf r})$ and
$\psi_a({\bf r})$ in the representation of occupation numbers
[13], it is required to choose a complete set of  wave functions
$\langle{\bf r} | \{Q\sigma \}, \{{\bf R}_i^c\}\rangle $
characterizing single-particle electronic states with the energy
$E(\{Q\})$
\begin{eqnarray}
\psi_e^+({\bf r}) =\sum_{\{Q\sigma\}}\langle \{Q\sigma \}, \{ {\bf
R}_i^c \} | {\bf r} \rangle a_{\{Q\sigma\}}^+, \qquad \psi_e({\bf
r}) =\sum_{\{Q\sigma\}}\langle {\bf r}| \{Q\sigma \}, \{ {\bf
R}_i^c \} \rangle a_{\{Q\sigma\}}, \label{F5}
\end{eqnarray}
in the external field of nuclei. Here $a_{\{Q\sigma\}}^+$ and
$a_{\{Q\sigma\}}$ are the creation and annihilation operators of
electrons in the state $\langle{\bf r} | \{Q\sigma \}, \{{\bf
R}_i^c \} \rangle $, $\{{Q\sigma} \}$ is the set of quantum
numbers, including the spin number $\sigma$ (hereafter, for
simplicity, spin indices are not considered). In the problem under
consideration the set of coordinates $\{{\bf R}_i^c\}$ of nuclei
plays the role of eigenvalue parameters. In this case,
single-particle electronic states in the external field of nuclei
are classified into two groups [17]. One (A) is localized
(subscript "loc") states characterized by exponential spatial
decrease with distance from localization centers. Among them are
the single-center bound states of electrons in atoms, two-center
ones in molecules, etc. Another group (B) is delocalized
(subscript "del") states in which the electron propagates
throughout the system. The typical example of delocalized states
are plane waves $\langle {\bf r}|{\bf q}\rangle$, $\langle {\bf
r}|{\bf q} \rangle = 1/\sqrt {V}\exp (i{\bf q r})$,
$\epsilon(q)=\frac{\hbar^2 q^2}{2m_e}$ which describe the behavior
of the free electron with momentum $\hbar {\bf q}$ and energy
$\epsilon(q)$. Thus, in the optimum case, the complete set of wave
functions for the representation (5) should describe both
localized and delocalized states of electrons. However, there are
many problems in this way. The first problem is "physical"; it is
associated with the fact that localized electronic states vary as
thermodynamic parameters of the CS are varied; this leads, in
particular, to the Mott effect [18], i.e., disappearance of
localized states. This means that the quantum-mechanical
consideration is insufficient for determination of localized
states; quantum statistics effects should also be considered. The
second problem, to the solution of which this paper is devoted, is
associated with the necessarily finite number of electronic states
localized on one center. By the example of the hydrogen atom
considered within quantum mechanics ignoring quantum statistics
effects [13], it is known that the number of bound electronic
state is infinite, which leads, in turn, to the divergence of not
only the statistical sum for an individual atom, but also to the
unlimited atomic size at any nonzero temperature [2]. Based on
combinatory reasons, E. Fermi [7] has shown that the introduction
of the finite size of atoms results in an exponential limitation
of the statistical sum. It is easy to see that the statistical sum
limitation to any maximum value of the principal quantum number
$n_{max}$ is also a simultaneous introduction of the finite size
of the atom. The absence of limitation, i.e, the case $n=\infty$,
is the case corresponding to an infinite-size atom, which is
meaningless (the size of the $n$-th orbit of the hydrogen atom is
$R_n=n^2 a_0$, where is the Bohr radius) [19]. Hence, it is
impossible to correctly use the "single-center" (in the sense of
the localization center) approximation in the description of
localized states. In this case, the transition to the
"single-center" approximation is meaningful when the condition
\begin{eqnarray}
<R_a>\ll <r>_c,\;\; <r>_c=\left(\frac{4 \pi n_c}{3}\right)^{-1/3},
\label{F6}
\end{eqnarray}
is satisfied, where $<R_a>$ is the "atom" size and $<r>_c$ is the
average distance between nuclei (see below).

The consideration of the quantum statistics effects (with the
opportunity to pass to the thermodynamic limit) means that
localized electronic states for a separated nucleus should be
considered taking into account that the CS, as a system of many
identical particles, contains many other electrons (electronic
states). However, the multiplicity of noninteracting electrons is
already formally taken into account in the used formalism of
secondary quantization (1), (2), (5).

The divergence directly follows from the definition of the
electron distribution function $f_e(E(\{Q \}))$ in states $\langle
{\bf r} | \{Q\}, \{{\bf R}_i^c\}\rangle$ with energy $E(\{Q\})$
\begin{eqnarray}
f_e(E) = \langle a_{\{Q \}}^+ a_{\{Q \}} \rangle ^{(GE)}= \left \{
\exp \left( \frac{E-\mu_e}{T} \right)+1 \right\}^{-1}, \qquad
\langle N_e \rangle ^{(GE)} = \sum_{\{Q\}}f_e (E), \label{F7}
\end{eqnarray}
when using energy levels of the hydrogen atom as energies. It
remains to hope that the "effective" consideration of the
electron--electron interaction while retaining the single-particle
consideration will make it possible to solve the problem of the
finite number of bound states in the "atom", i.e., in the
"single-center" approximation for wave functions (it is clear that
the "multicenter" problem for localized electronic states is in
principle unsolvable),

\begin{eqnarray}
\langle {\bf r}|\{loc\}, \{{\bf R}_i^c\}\rangle \to \langle {\bf
r}- {\bf R}_i^c |n \rangle \equiv \varphi_n ({\bf r}- {\bf
R}_i^c), \qquad E(n, \{loc\}) \to E_n .
 \label{F8}
\end{eqnarray}

In this case, the "atom" size in the "single-center"
approximation, taking into account (7) and (8), can be defined for
quantum statistics as follows
\begin{eqnarray}
<R_a>^2=\int r^2 \sum_n f_e (E_n)|\varphi_n(r)|^2 d{\bf r}.
 \label{F9}
\end{eqnarray}

We emphasize that such quasiparticles - "atoms" are not identical
to atoms commonly used in quantum mechanics. The difference is
caused by the fact that any of one of a very large (infinite in
the thermodynamic limit) number of electrons can be found in the
corresponding electronic state of the "atom" due to the identity,
which is reflected in the statistical description of such a
system. Thus, electronic states in "atoms" will depend on the
thermodynamic parameters, i.e., the temperature $T$ and average
density of the number of electrons $n_e^{(0)}$ (or on the chemical
potential $\mu_e$) [2].

We note that electronic states are considered in the random field
of nuclei. In this case, the role of quantum numbers for both
"localized" (with discrete energy spectrum) and "delocalized"
(with continuous energy spectrum) electronic states is played by
the energy itself of the corresponding state and the spin quantum
number [20]. For "localized" states, this set of quantum numbers
can be complemented by coordinates of localization centers
(nuclei) $\{{\bf R}_i^c\}$ [20], since electronic states with this
energy can become localized at different points of the system. In
this sense, we can speak of degeneracy with respect to the
quantities $\{{\bf R}_i^c\}$ which, therefore, can be considered
as "quantum" numbers. In this case, the term "quantum number"
should be understood in a certain conditional meaning: the vectors
$\{{\bf R}_i^c\}$ and their components are not eigenvalues of any
operator having an effect on "electronic" variables.

\section{Finite atom in Hartree-Fock approximation}

Further, let us consider the system of electrons in the external
field of nuclei within the self-consistent Hartree-Fock
approximation for quantum statistics [15]. Then the complete set
of wave functions $\langle {\bf r} | \{Q\}, \{{\bf
R}_i^c\}\rangle$ and the corresponding energy spectrum $E(\{Q\})$
of the multicenter problem are determined from the set of
equations
\begin{eqnarray}
\left\{- \frac{\hbar^2\nabla^2}{2m_e}+ \int [u_{ec}( {\bf r}- {\bf
r}_1)n_c({\bf r}_1)+ u_{ee}( {\bf r}- {\bf r}_1)
\langle n_e ({\bf r}_1) \rangle ] d {\bf r}_1 \right \} \langle {\bf r}| \{{Q}\}\rangle -\nonumber\\
\int u_{ee}( {\bf r}- {\bf r}_1) \langle n_e ( {\bf r},{\bf r}_1)
\rangle \langle {\bf r}_1|\{{Q}\} \rangle d {\bf r}_1 = E
(\{{Q}\})\langle {\bf r}| \{{Q}\} \rangle , \label{F10}
\end{eqnarray}

In (10), the term with function $\langle n_e ({\bf r}) \rangle =
\langle n_e ({\bf r}, {\bf r}_1) \rangle |_{{\bf r} = {\bf r}_1}$
corresponds to the Hartree approximation. The term with function
$\langle n_e ({\bf r}, {\bf r}_1) \rangle = \sum_{\{Q \}}\langle
{\bf r} | \{Q \} \rangle \langle \{Q\}|{\bf r}_1\rangle
f_e(E(\{{Q}\}))$ in (10) corresponds to the Fock approximation.
Nonlocal structure of the Fock term in (10) significantly
complicates the problem. However, its consideration is
conceptually necessary. First, in the Hartree approximation, the
states $\langle {\bf r}|\{Q\}, \{{\bf R}_i^c\}\rangle$ are not
orthogonal to each other [13]. Second, the consideration of the
Fock approximation allows elimination of the so-called
"self-action". As it is easy to check from (10), the state
$\langle {\bf r}|\{Q\}\rangle$ in the Hartree-Fock approximation
is determined only by other states, in contrast to the Hartree
approximation [13]. The set of equations (10) includes not only
localized, but also delocalized states, to describe which the
"single-center" approximation is inapplicable. Taking into account
that we are interested in justification of the finiteness of the
number of localized states, let us simplify the set of equations
(10) under the assumption that the number of delocalized states is
very small. This can be done using the properties of the
Fermi-Dirac distribution (7) for the function $f_e(E)$ in (10).
Let us perform the further consideration without loss of
generality immediately for the case of hydrogen, $z_c = 1, \langle
N_e \rangle^{(GE)} =N_c$. Let us suppose that for the energy $E_0$
of the ground single-particle localized state, energies $E_{exc}$
of each of other states, including delocalized states, and the
chemical potential $\mu_e$ of electrons, the following relations
are valid
\begin{eqnarray}
E_0 < 0, \qquad | E_0| \gg T, \qquad  | E_0- E_{exc}| \gg  T,
\qquad  E_0 < \mu_e < 0.
 \label{F11}
\end{eqnarray}

Then, according to (7),

\begin{eqnarray}
f_e (E_0) \cong 1 - \exp \left(\frac{ E_0-\mu_e}{T} \right) \cong
1,\qquad  f_e (E_{exc}) \cong \exp \frac{\mu_e -E_{exc}}{T}\ll 1.
\label{F12}
\end{eqnarray}

We further take into account that the number of single-particle
localized states is proportional to the number of localization
centers; in the "single-center" approximation, it is proportional
to the number of nuclei $N_c$. We also assume that the delocalized
states are close to corresponding plane waves. Then, with required
accuracy, we can consider that the energy levels of delocalized
states are defined by the energy $\epsilon(q)$ of the free
electron, $E(\{del\}) \to \epsilon(q)$. Thus, taking into account
\begin{eqnarray}
\langle N_e \rangle^{(GE)}= \sum_{\{Q\}} f_e (E(\{{Q}\})) \to N_c
\sum_n f_e(E_n)+ \sum_q f_e (\epsilon(q)), \label{F13}
\end{eqnarray}
and (7)-(12), we find the equation for determining the chemical
potential $\mu_e$ of electrons in the "single-center"
approximation for localized states,
\begin{eqnarray}
\langle N_e \rangle^{(GE)}= N_c \left \{1-\exp \left( \frac{ E_0-
\mu_e }{T} \right) +\sum_{n\ne 0}\exp \left( \frac{ \mu_e -
E_n}{T}\right) \right\} + \sum_q \exp \left(
\frac{\mu_e-\epsilon(q)}{T} \right) . \label{F14}
\end{eqnarray}

Then, taking into account (11) and the equality
\begin{eqnarray}
\sum_q \exp \left( \frac{\mu_e-\epsilon(q)}{T} \right) = V \int
\exp \left( \frac{\mu_e-\epsilon(q)}{T} \right)
\frac{d^3q}{(2\pi)^3}= \frac{V}{\Lambda^3}\exp \left( \frac{\mu_e
}{T}\right), \qquad \Lambda = \left( \frac{2\pi\hbar^2}{m_eT}
\right)^{\frac12},  \label{F15}
\end{eqnarray}

from (14), we find
\begin{eqnarray}
\mu_e=E_0 - \frac T2 \ln \left \{ \sum_{n \ne 0}  \exp \left(
\frac{E_0-E_n}{T} \right) + \frac{1}{n_e\Lambda^3} \exp \left(
\frac{E_0}{T}\right) \right\}. \label{F16}
\end{eqnarray}

Thus, conditions (11) for the chemical potential are satisfied
when
\begin{eqnarray}
\sum_{n \ne 0} \exp \left( \frac{E_0-E_n}{T} \right) +
\frac{1}{n_e \Lambda^3}\exp \left( \frac{E_0}{T} \right) < 1 .
\label{F17}
\end{eqnarray}

Taking into account (11), inequality (17) is valid for a limited
number of localized states. According to (11), (12), (14), the
contribution of delocalized states in the self-consistent
Hartree-Fock approximation (10) can be neglected in a very wide
range of thermodynamic parameters satisfying the condition
$\sum_{n \ne 0} \exp \left (-\frac {E_n} {T} \right) \gg \frac {1}
{n_e\Lambda^3}$. When determining localized states and the
corresponding energy spectrum, we can consider with required
accuracy that the ground state is "populated", $f_e (E_0) \cong 1$
and all excited states are "free", $f_e (E_{exc}) \cong 0$. Thus,
when relations (11), (16), (17) are valid, the set of equations of
the self-consistent Hartree-Fock approximation (10) can be written
for localized states in the "single-center" approximation (8) (the
nucleus with charge $z_c=1$ is at the coordinate origin),
\begin{eqnarray}
\left \{ - \frac{\hbar^2\nabla^2}{2m_e}- \frac{e^2}{r}+ \int
u_{ee}( {\bf r}- {\bf r}_1) | \varphi_0( r) |^2 d {\bf r}_1\right
\} \varphi_n( r) - \nonumber\\
\int  u_{ee}( {\bf r}- {\bf r}_1)\varphi_0( r) \varphi_0^*( r_1)
\varphi_n( r_1)d {\bf r}_1 = E_n \varphi_n (r), \label{F18}
\end{eqnarray}

It immediately follows from (18) that the wave function $\varphi_0
({\bf r})$ and the ground state energy $E_0$ are identical to the
similar quantities for the electron ground state in the hydrogen
atom [13], $\varphi_0(r) = (\pi a_0^3)^{-\frac 12} \exp
\left(-\frac {r}{a_0}\right), \quad E_0 =-\frac {m_ee^4}
{\hbar^2}$, where $a_0$ is the Bohr radius. The "excited"
localized states $(n> 1)$ are characterized by the energies $E_n$
and differ from corresponding states in the hydrogen atom, which
makes it possible, in particular, to remove the known "Coulomb
degeneracy" of energy levels [13]. Taking into account the
explicit form of the Fourier components of the Coulomb interaction
potential (2), we can calculate the Hartree term, i.e., the last
braced term in (18),
\begin{eqnarray}
u_H(r)= \int u_{ee}( {\bf r}- {\bf r}_1) |\varphi_0({\bf r}) |^2 d
{\bf r}_1 = \frac{e^2}{r} - \frac{e^2}{r} \left( 1+ \frac{r}{a_0}
\right) \exp \left( -\frac{2r}{a_0}  \right) . \label{F19}
\end{eqnarray}

Further analytical calculations seem impossible. As in
quantum-mechanical calculations, in view of the nonlocality of the
Fock term, equations (18), taking into account (19), are solved
using iterations [13,14]. However, by analogy with the states in
the hydrogen atom [13], we can assume that the limit relations
\begin{eqnarray}
\lim_{r \to 0}\varphi_n(r)=c_n < \infty , \qquad \varphi_n( r\to
\infty ) \sim \exp \left( - \frac{a_nr}{ a_0}  \right), \qquad 0<
a_n < 1. \label {F20}
\end{eqnarray}
are valid for functions $\varphi_n (r) (n > 1)$.

In this case, the Fock term in (18) and (19) can be written in the
"local" form
\begin{eqnarray}
\int  u_{ee}( {\bf r}- {\bf r}_1)  \varphi_0( r) \varphi_0^*( r_1)
\varphi_n( r_1)d {\bf r}_1 = u_F^n(r)\varphi_n( r), \label{F21}
\end{eqnarray}

\begin{eqnarray}
u_F^n(r)=\varphi_0( r) \varphi_n^{-1}( r)  \int  u_{ee}(q)\exp
(i{\bf q r}) \phi_{0n}(q) \frac{d^3q}{(2\pi)^3} , \label{F22}
\end{eqnarray}

\begin{eqnarray}
\phi_{0n}(q)= \int  \varphi_0( r) \varphi_n( r)  \exp (-i{\bf q
r}) d {\bf r}.
 \label{F23}
\end{eqnarray}

According to (20) and (23),
\begin{eqnarray}
\lim_{q \to\infty} \phi_{0n}(q)=0, \quad \lim_{q \to 0}
\phi_{0n}(q)= \phi_{0n}(q=0)= \int  \varphi_0( r) \varphi_n( r) d
{\bf r}=0.
 \label{F24}
\end{eqnarray}
due to the orthogonality of the functions $\varphi_0 (r)$ and
$\varphi_n (r)$. Then it immediately follows from (21)-(24) that
\begin{eqnarray}
\lim_{r \to 0}u_F^n(r)=u_n = const, \qquad u_F^n(r \to \infty )
\sim \exp \left( - \frac{\gamma_nr}{a_0}  \right), \qquad \gamma_n
> 0.
 \label{F25}
\end{eqnarray}

Hence, we come to the conclusion that the function $\varphi_n(r)$
is defined as the wave function satisfying the Schrodinger
equation,
\begin{eqnarray}
\left( - \frac{\hbar^2\nabla^2}{2m_e} +U_n(r) \right)
\varphi_n(r)=E_n\varphi_n (r), \nonumber\\
U_n (r \to 0)=-\frac{e^2}{r}, \qquad  U_n (r \to \infty ) \sim
\exp \left(- \frac{\gamma_nr}{a_0} \right).
 \label{F26}
\end{eqnarray}

Thus, the initial assumptions (20) are valid. Furthermore, the
asymptotic behavior of the potential $U_n(r)$ is similar to the
behavior of the known Yukawa potential [21]. As is known, at
certain parameters of the Yukawa potential, it will not contain
the bound (localized) state [22]. As applied to the problem under
consideration, this means that the number of localized states in
the hydrogen "atom" is limited.

Using the results obtained above, it is easy to see that the free
energy $F^{CS}$ (3) of the Coulomb system under consideration
equals
\begin{eqnarray}
F^{CS}=-N_c T \left\{1+\ln \left[\left(\frac{m_c T}{2\pi
\hbar^2}\right)^{3/2}\frac{f_e(T)}{n_c}\right]\right\},\;\;\;
f_e(T)=\exp\left(-\frac{E_0}{T}\right) \label{F27}
\end{eqnarray}
if the conditions (6), (11), (16), (17) are valid.

The relation (27) corresponds the free energy of the ideal gas of
new quasiparticles - "atoms", in which electron is in the ground
state (see, e.g., [16]).

Summarizing the above consideration, it can be assert that the
statistical sum and the hydrogen "atom" size are finite when using
the self-consistent quantum-statistical Hartree-Fock
approximation. In this case, in the above approximation (11) for
the chemical potential of electrons, the "atom" size (9) is
independent of the density of the number of particles and of
temperature in the system and is defined only by the Bohr radius
in a wide range of thermodynamic parameters. For low temperature
the developed approach provides the rigorous derivation of the
atomic free energy (27), written earlier on the basis of physical
reasons. On the basis of the above derivation, it is evident that
the finiteness of the number of localized states leads to the
finite free energy of the CS at arbitrary temperature.

\section*{Acknowledgments}

The authors thank W. Ebeling and A.L. Khomkin for the useful
discussions. This study was supported by the Netherlands
Organization for Scientific Research (NWO), project no.
047.017.2006.007, and the Russian Foundation for Basic Research,
projects no. 07-02-01464-a and no. 10-02-90418-Ukr-a.

\end{document}